# Can we build a conscious machine?

Dorian Aur[1]

*Abstract:* The underlying physiological mechanisms of generating conscious states are still unknown. To make progress on the problem of consciousness, we will need to experimentally design a system that evolves in a similar way our brains do. Recent experimental data show that the multiscale nature of the evolving human brain can be implemented by reprogramming human cells. A hybrid system can be designed to include an evolving brain equipped with digital computers that maintain homeostasis and provide the right amount of nutrients and oxygen for the brain growth. Shaping the structure of the evolving brain will be progressively achieved by controlling spatial organization of various types of cells. Following a specific program, the evolving brain can be trained using substitutional reality to learn and experience live scenes. We already know from neuroelectrodynamics that meaningful information in the brain is electrically (wirelessly) read out and written fast in neurons and synapses at the molecular (protein) level during the generation of action potentials and synaptic activities. Since with training, meaningful information accumulates and is electrically integrated in the brain, one can predict, that this gradual process of training will trigger a tipping point for conscious experience to emerge in the hybrid system.

**Keywords:** *brain; consciousness; neuron; synapses; nanoneuroscience; neural plasticity; neuroelectrodynamics.*

**Introduction**

For a long time philosophers, neuroscientists and cognitive psychologists have speculated about the nature of consciousness. Science has made amazing advances, however, the nature of conscious processes still remains a mystery and a controversial topic (Searle,1980; Penrose, 1994; Hameroff & Penrose, 1996; Bennett, 1997; Chalmers, 1997; Ford, 2000; Arbib, 2001; Dehaene and Naccache, 2001;Dennett, 2002; Baars, 2002; Lehar, 2002; Josephson, 2002; Josephson, 2003; Spivey, 2007; Nunez, 2010; Vimal, 2009; Acacio de Barros, & Suppes, 2009; Damasio, 2012; Dehaene and Changeux, 2011; Koch, 2012; Pereira & Lehmann, 2013;Searle, 2013; Cacha and Poznanski, 2014; Hameroff et al., 2014 ; Cornelis & Coop, 2014 ).

---

[1] Email : DorianAur@gmail.com

Current attempts to replicate the human brain and consciousness on digital computers are not based on a clear assessment of brain complexity. As recently predicted, these projects have little chance of success (Cattell and Parker, 2012; Alivisatos et al., 2013; Fields, 2013; Sejnowski & Delbruck, 2012).

Many controversies in neuroscience are generated by unreliable hypotheses. The idea that coding of sensory input is performed in the brain by the timing of spikes, is a result of ill-defined intuition. The timing of spikes solely provides evidence that neurons respond to sensory inputs. Temporal patterns do not necessarily tell what kind of information is stored or communicated (Aur and Jog, 2010).

In electrophysiology (neuroscience), action potentials are described as a change in the membrane potential, however in physics or electrical engineering they will always remain non-stereotyped electrical waves (Aur et al., 2005; Aur et al., 2011). Even one records only the envelope of an action potential, every spike carries imperceptible endogenous waves and nonlinear vibrations generated by molecular structures (Aur and Jog, 2010; Woolf et al., 2009; Cifra et al., 2011; Wu, 2005). Indeed, for us these rhythms are imperceptible, however such waves do not seem to be ignored by surrounding neurons and synapses.

Always in the brain the electric field penetrates biological structure and during an action potential this variable electric field carries molecular vibrations (Fraser & Frey 1968, Martí & Bishop, 1993, Park and Boxer, 2002). Higher wave frequencies are generated by smaller structures embedded inside cells. Experimentally, these micron wavelengths are part of a large electromagnetic spectrum generated during action potential propagation (Fraser & Frey, 1968). Since meaningful information is stored at a molecular (protein) level in neurons and synapses, these terahertz waves are highly important (Born et al., 2009; Cifra et al., 2011). In this case, the carrier frequency can have a much lower frequency than the 'modulating' waveform. As a result, meaningful information embedded within molecular structure (e.g. proteins) is carried out by electrical waves. An action potential represents the moment of "reading out" meaningful information from molecular structures. During a spike event, meaningful information can also be "written" at a molecular scale in neurons and synapses since subcellular changes in the pattern of gene expression can be easily triggered by external events (Ivanova et al., 2011). A similar phenomenon occurs during synaptic activities when a flow of ions generate molecular vibrations and electric waves. The smaller the structure, the higher the generated rhythm of transformation can be. Therefore, meaningful information that was encoded (written) within neurons and synapses at a molecular level is transmitted synaptically and non-synaptically (wirelessly in both cases) during action potential propagation and all this meaningful information is electrically integrated in the brain (Aur, 2012b; Nunez & Srinivasan, 2006; Cifra et al., 2011; Landfield & Thibault, 2001; Aur and Jog, 2010;).

Almost all mathematical models highly simplify these phenomena. Only two waves, (forward and backward waves) are included in the FitzHugh-Nagumo model. In reality, the action potential is composed from a large number of waves with smaller amplitudes generated by molecular vibrations. The occurrence of meaningful electrical patterns within the cell during an action potential generation can be easily explained as a physical process of wave interference in which at least two waves superpose in space (Aur and Jog, 2010; Acacio de Barros, & Suppes,2009). Contextual wave interference explains the presence of stronger nonlinear vibrations within certain parts of recorded cells that provide fragments of the engram (Aur, 2012b). Importantly, it also highlights the importance of space at larger scales inside the brain where neurons are part of the most powerful computational device known.

Physics offers simple explanations for such processes. In other words, matter within the neuron and synapses is represented by their meaningful waves and these waves carry meaningful information. What we have called matter inside neurons (e.g. proteins) is also energy, "whose vibration has been so lowered as to be perceptible to the senses" (A. Einstein). The method to interface TMS with EEG (Grau et al., 2014) demonstrates beyond doubt that inside the brain meaningful information can be consciously read out (decoded), written (encoded) and transferred using the electromagnetic spectrum.

Surrounded by polar water, molecules ensembles of proteins also generate vibrations and behave as crystal lattices which can trap ions and electrons. In addition to neurotransmitters and hormones, the presence of water channels (aquaporins) and stretching vibrations on femtosecond timescales allow a fast redistribution of energy enhanced by polar water molecules which can also carry structural and dynamical information (Chen et al.,2010; Chopra and Levitt, 2011). Recently, a group led by Nobel laureate L. Montagnier has detected low-frequency electromagnetic waves emitted by the DNA of bacteria and viruses in low dilution solutions (Montagnier et al., 2011). Neither the contribution of electric fields nor structural dynamics of polar water molecules can be easily replicated on digital computers to build reliable models. In addition, understanding the functional role of bio-photon emission and absorption in living systems especially in proteins is still debated (Pang, 2012, Bokkon et al., 2013).

From computational models presented in neuroelectrodynamics(Aur and Jog, 2010; Aur, 2012a) one can predict that experimentally a machine which will physically interact with matter in a similar way our brains do will allow consciousness to emerge. To test this hypothesis a hybrid model that includes an evolving human brain and a digital computer system has to be built (**Figure 1**).

**Background and Significance**

Recently, the Japanese Nobel Prize-winning Shinya Yamanaka has shown that the specialization of cells is reversible. Mature cells can be easily converted into stem cells (Takahashi and Yamanaka, 2006). The presence of electromagnetic fields can also mediate nuclear reprogramming (Baek et al., 2014)

In addition, the experiment of Lancaster et al., 2013 can be used as a model to build evolving brains with functional neurons that survive indefinitely. Only after few days the organoids can develop functional neurons which later differentiate into rudimentary brain regions (prefrontal cortex, occipital lobe, hippocampus and retina). However, the above experiment didn't form the exact same regions as in a normal brain and the evolution stopped after 2 months.

Having a model that reproduces the development of the human brain is critical to understand the roots of consciousness. Since global functions do not occur in brain slices or dishes of neurons the importance of the whole (intact) brain has been repeatedly stressed by Steriade (Steriade, 2001)

Current progress in growing cerebral organoids has open new possibilities. Embryoid bodies can be grown either from natural stem cells, or from induced pluripotent cells that have the potential to generate all cell types (Discher et al., 2009; Pasca et al.,2014).

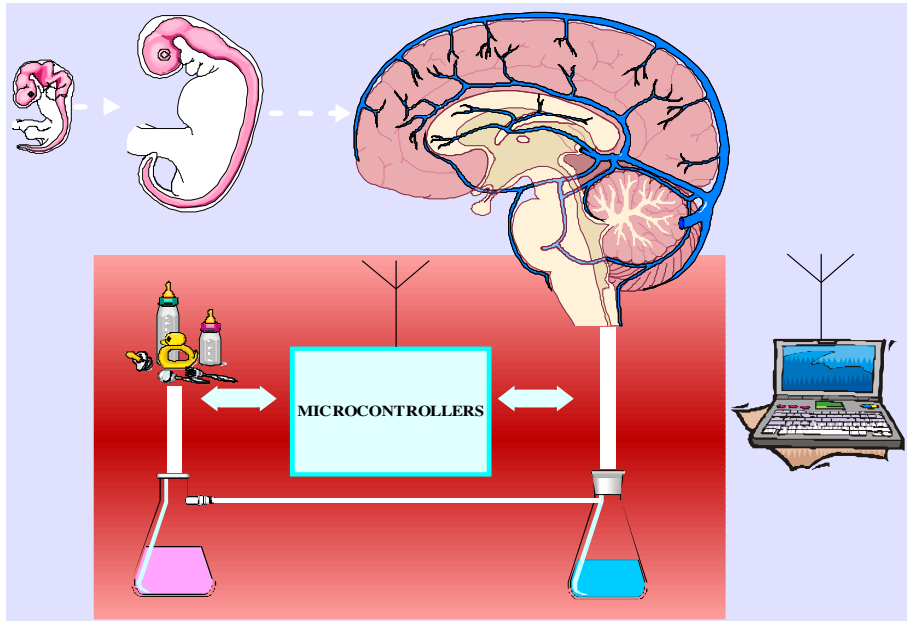

**Figure 1:** Schematic representation of evolution from cerebral organoids or pluripotent cells. Equipped with nanosensors, microcontrollers, analog to digital convertors (ADCs) and wireless communication protocol, the digital computer maintains homeostasis and provides the right amount of nutrients and oxygen for human brain growth.

In electrochemical interfaces, soluble carbon nanotubes (CNT) can efficiently conduct electrical current and create physical support for brain development. CNT can provide the required coupling with amino acids and bioactive peptide (Bianco and Prato, 2003) and promote neuron differentiation (Chao et al., 2009). Using dielectrophoresis, cells can be selectively manipulated based on their phenotype while structural changes can be monitored by acquiring cross-over frequencies (Gascoyne, 2013). Importantly, this project is required to bypass limitations imposed by animal models in case of schizophrenia, autism and many other neurological disorders.

**Materials and Methods**

To make progress on the problem of consciousness, one needs to build a system that evolves in a similar way our brains do.

*The first phase will require* growing a full size brain either from natural stem cells or from induced pluripotent cells. Providing nutrients, oxygen and environmental interaction is needed to shape the structure of the evolving human brain and control spatial organization of cells .

*The second phase will* create a virtual world in which the evolving human brain can be trained to learn and experience live scenes following a specific gradual program. It is likely that after training the hybrid system will be able to mimic human behavior in the 'real' world and trigger conscious experience.

The first phase will require developing a system and technology to grow full size human brains. The entire evolution of the brain will be regulated using a brain computer interface (**Figure 1**). Equipped with microcontrollers and different nanosensors (Cui et al., 2001), the digital computer will obtain real-time information regarding the state of the evolving brain and detect the need of neurotrophic factors, nutrients and oxygen (Lamas et al., 2014). This phase will allow various organoids to self-assemble and organize into discrete, interdependent domains (Bond et al., 2005; Evans et al., 2005). Different ways to deliver nutrients, oxygen, and achieve spatial and temporal control of living tissue by manipulating molecular and genetic technology can be explored (Delcea et al., 2011; Lewandowski, et al., 2013; Takebe et al., 2013; Deisseroth and Schnitzer, 2013; Wickner and Schekman, 2005). Dielectrophoretic actuation will be used for cell manipulation to shape the evolving 3D structure (Pethig et al., 2010; Reyes, 2013; Velugotla et al., 2012). In addition, carbon nanotubes will provide the physical support for brain development. They can be used to create conductive structures to perform bidirectional communication between the evolving brain and computers. This will allow monitoring the evolution of neurons, glial cells, brain size, achieving characteristics similar to real brain by delivering neurotrophic factors and engineering all brain structures. Once the brain regions attain the development of a new born baby, the training phase using substitutional reality can be started.

The absence of blood vessels can generate difficulties to deliver nutrients, oxygen and may stop the evolution. In addition to other artificial techniques, the protocol described by Takebe et al., 2013 to build a three-dimensional vascularized organ will be explored. Nanosensor devices (e.g. carbon nanotubes) can be used to monitor electrical activity, oxygen, or chemical levels and provide reliable details regarding the path of evolution which can be controlled by the digital computer.

*The second phase will use computer technology to create a virtual world and provide accelerated training.* Substitutional reality will enhance learning, the evolving brain will be able to mimic human behavior in the real world.

The main goal of this step would be to build bidirectional communication between the evolving brain and the computer to create a virtual world and enhance learning (**Figure 2**). One can read and interpret the information processed in the evolving brain by using data recorded from different nanosensors (Pasley et al., 2012; Akemann et al., 2013; Bernstein et al., 2011; Bargmann & Newsome, 2014; McClelland et al. 2013). The entire model can be schematically conceptualized as an interactive training system that shapes the evolution of the brain based on natural language and visual information (Heinrich et al., 2014). Initially, the computer will generate a sensorimotor flux that a baby is confronted with. This early phase will allow cerebral organoids to self-assemble into a functioning system. Gradually, the computer will generate naturalistic and contextually rich scenarios dependent on the evolutionary stage which will allow the development of behavioral responses in the virtual world analogous to those that occur in the real world. The evolving brain will experience live scenes and learn from interactions (Bohil, et al., 2011; Suzuki et al., 2012). This experiment will be used to further study pathophysiological mechanisms of induced neurological disorders and provide insights regarding potential treatments (Jain, 2001).

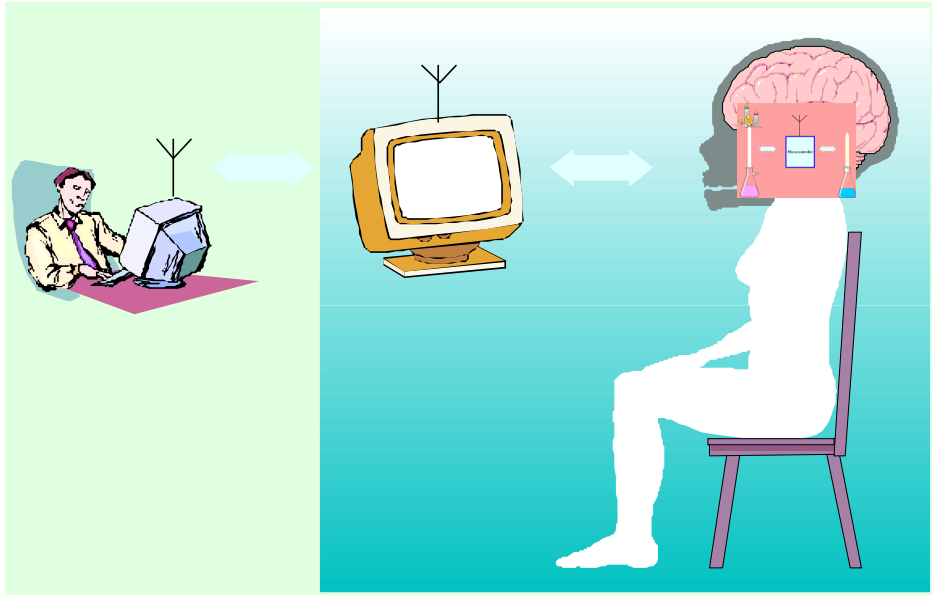

**Figure 2**: Substitutional reality is a non-expensive technique to train the evolving brain. The computer creates a virtual world in which the evolving brain is trained to learn from interaction and perform human behavior. Bidirectional wireless communication between the brain and digital computers will serve for interactive training.

Acquiring cognitive abilities and other features unique to the human nature is a challenge (Pinker, 2003). Substitutional reality is a non-expensive method to train the evolving brain and can be used to study the formation of cognitive functions and psychiatric disorders in an evolving brain setup. Since virtual environments may not be always as richly textured as the real world, a combination between substitutional reality and natural environment may be required. Bidirectional communication will become functional once the evolving brain will be able to control its environment using the computer for basic needs. Once the evolving brain will be able to 'move' into the virtual space, the system will acquire cognitive abilities and other features unique to the human nature. With interdisciplinary collaboration this path of reprogramming human cells is far more plausible than any competing scenarios of mapping the entire human brain or instantiate consciousness on digital computers.

This project represents an interdisciplinary approach to reliable design conscious machines and does not require a full understanding of consciousness. Animal models may also experience some form of consciousness and to avoid any ethical issues such experiments can be initially performed using nonhuman cells.

**Conclusion**

This endeavor will be equivalent to the Moon landing project or the discovery of the Higgs particle. Having included *evolutionary relationships*, the platform can achieve far more than competing projects (e.g. Blue Brain, Human Brain Project). Since from a nanoscale level almost everything can be manipulated at will, the project will provide a clear alternative to understand how the brain functions from a molecular level. With various types of nanosensors any change of the evolving brain can be well monitored, recorded and then simulated on digital computers. Such project will reduce the technological barrier to understand the evolution of the human brain, neurodevelopmental processes, different neurological diseases and will represent a reliable platform for treatment in severe neurological disorders, drug testing, gene therapy and nano experiments. In addition, it will significantly reduce the excessive costs and inefficiency associated with animal testing providing less dependency on animal models. The new platform will help to expand natural intelligence, design new types of (humanoid) robots that use natural language as forms of communication, and it can be used in environments where humans can't survive and will generate high economic impact. Finally, it will reveal how we do integrate what we see, how feelings, emotions are triggered. It will provide a deep view of human nature and shed light on the nature of consciousness.

Currently, an active preservation of the human brain with little or no loss of information is science fiction; however, this platform will technologically open a new path. In a set-up where the real human brain can accommodate to work directly with digital computers which are able to maintain physiological homeostasis, even "afterlife" becomes possible. Such platform can create reliable premises to avoid brain death or even replace cryopreservation in order to achieve 'immortality'. I believe that this kind of hybrid, evolving system is the future. It's a dream that can change the way we treat brain diseases or understand the nature of human consciousness.

*Acknowledgments*: The author would like to thank Brian Josephson, Michael J. Spivey, Michal Cifra, Allan D. Coop , Thibault Olivier, Bernard J. Baars, Mandar S. Jog, Paul L. Nunez, Alfredo J. Pereira, Roman Poznanski for excellent suggestions.

*Note: The author would like to collaborate with any entity, academic or private laboratory, research institute, foundation, consortium or private investors to build the first prototype.*

**Supplementary Material: Selected Questions and Answers**

*Some ideas are irrationally perceived as dangerous and protective mechanisms, usually involving arguments that would fall apart under close examination, are brought up to avoid confronting the possibility that they may be of value. -B. Josephson*

- **This is science-fiction, I've read a very similar novel once**

Indeed, this is the general perception. Recently, Takahashi and Yamanaka have experimentally proved that the specialization of cells is reversible (no theory predicted this outcome) and this result opens fast new possibilities. Jules Verne was always considered a science fiction writer. We did even better than he predicted (Twenty Thousand Leagues Under the Sea, and Around the World in Eighty Days). Jules Verne is still considered a science fiction novelist. What would write Jules Verne today?

- **We are doing early science and "conscious machines" seem premature**

Everyone thinks that today we are very far away from building anything remotely resembling a conscious machine. This perception might change once one understands the limits of digital computers and that our brain uses a different; more powerful "form of computation". Indeed, *"conscious machines" are science fiction if one solely uses digital computers*. In this field of conscious machines the capabilities of digital systems are limited. Lately, many leaders of AI research still dream to build such systems, however far less than ever before. After decades of major problems in artificial intelligence building conscious machines became science fiction. We can solve these issues by using hybrid systems that include evolving biological structures. Once one understands that such endeavor is possible, the field of hybrid (intelligent) systems will explode. *Whenever there has been progress, there have been influential thinkers who denied that it was genuine, that it was desirable, or even that the concept was meaningful.- David Deutsch*

- **This experiment proposal of using human stem cells is unethical. How do you plan to solve this issue?**

"Brain in a vat" experiments are currently performed all over in the U.S. and Europe. Since animal models may also experience some form of consciousness one can use them to avoid any ethical issues.

- **The paper doesn't address the challenge of conscious machines it's just a bioengineering proposal to build a brain, that's not a machine**

Several references cited in this paper present the challenges of understanding consciousness; today we have far more questions than answers. Indeed, the paper highlights a series of interdisciplinary techniques that can be used to bioengineer conscious states in an artificial brain. The outcome fits the definition of a machine that can be truly alive.

- **Substitutional reality will not work, the "child" is clearly being raised in a deprived environment**

A combination between substitutional reality and real environment can be used to overcome this problem.

- **Having an actual simulation data is required to prove the concept, (e.g. resonance in microtubules, within an axon)**

The simulation on digital computers can be done, however it is unlikely that it will generate any form of consciousness. It's like passing a current that heats a wire. At a certain point in time with an increase in current intensity, the "light" becomes visible. Similarly, training, accumulating meaningful information will reshape the evolving brain structure. However, one can simulate over and over this process of 'heating a wire' (see the Blue Brain, Human Brain Project), no light, no emergent consciousness on digital computers.

- **However how will this be done? How long it will take? What is the cost?**

The total cost of finding the Higgs' boson was about $13.25 billion dollars and it took almost 40 years to reach an agreement in physics. Neuroscience is far more divided today than was physics 50 years ago. However, the interest on developing (conscious) intelligent machines or testing new drugs on a reliable human brain model may lead to a radical change. This interdisciplinary project may bring everyone together.

The entire experiment can start from any academic lab able to generate neurons in vitro and can continue in the backyard of many companies that have already developed techniques for substitutional reality e.g. Google, Microsoft, IBM. This kind of experiments can be carried on almost everywhere, in the US, Germany, UK, France, China, Korea, Russia, Australia, or even in Switzerland where the Human Brain Project has started. In addition, any pharmaceutical company (Novartis, Pfizer,..) interested to test drugs in real brains, not in an 'artificial' culture may also start such project. With interdisciplinary/international collaboration probably the first machine can be built in less than 5 years. The entire cost would be far less than 1 billion euro, see the initial cost of the Human Brain Project http://www.livescience.com/46678-scientists-criticize-human-brain-project.html

If Jules Verne would have lived today he would probably predicted 'miniature brains' in every cell phone, in every car and even in every house. In order to replace the real driver Google's driverless cars is bristled with very expensive sensors (e.g. laser eyes). It's a huge market for (conscious) hybrid machines and this project will provide competitive advantages for anyone that does research in this field.

- **It doesn't make sense to replicate a "living" human brain just to understand consciousness**

Understanding the nature of consciousness is just one goal. Testing other hypotheses including extrasensorial perception (ESP) phenomena, testing new drugs for the brain or carry any other experiments that cannot be performed on humans is highly important. The brain is built on different computing principles. Building intelligent (conscious) systems that can talk, move and solve problems like us will be the goal of many companies in the future.

- **In the early 1990's they could maintain a whole rat brain in vitro indefinitely. It was not known what the internal state of the brain was. Your experiment is similar**

Biological homeostasis is just a small part. Growing a brain and then gradually building meaningful communication with a computer is a different issue. In addition, this project is not solely about building a brain machine interface and definitely it cannot be reduced to neurons in a dish.

- **What about intelligence, creativity? How smart it will be?**

Quantitative genetic studies have shown beyond doubt that intelligence is highly inheritable, so the genetic material is important for intelligence and probably for creativity.

- **The human brain has about 100 billion neurons. How many neurons are needed?**

The entire project can begin with some neurons, sensors, interfaces, and a few computers. Maintaining homeostasis and providing the right amount of nutrients and oxygen is critical. The project requires extreme experimentation and it is likely that some forms of consciousness can be instantiated with fewer neurons. Since space in the brain is important to generate contextual wave interference, maintaining relative positions between cells is highly important.

- **Your paper does not cover every aspect neither regarding the brain growth nor the interface or training?**

Indeed, it doesn't cover 'every aspect', however many references are included. They provide many details so this paper can be transformed into a book. In addition, we will learn many things trying to build the system. At this stage probably "every aspect" is too much.

- **I plan to use billions of spiking neurons and with new memristor technologies we can build conscious machines**

Artificial spiking neurons and memristor technologies can be used to design adaptive systems. Alone they cannot build conscious machines. However, if evolving biological structures are included conscious (intelligent) machines can be built.

- **With such hybrid systems are we going to loose jobs in neuroscience, neural computation....?**

The perspective of finding jobs in these fields, in computer science, electronics, nanotechnology, machine learning will dramatically increase and compensate for any loss in other fields.

- **Can these systems be used to win wars?**

For several decades we understood the damage generated by nuclear attacks and successfully avoided the problem. We can also wisely use these hybrid systems.

- **Google has enough resources, they will build fast the first prototype and will not acknowledge your contribution? Does this bother you?**

I'll be grateful if Google or any other major company succeeds, their victory will be ours too. It will change forever the world we live in. However, sometimes successful companies fail to adapt. There are seized opportunities and missed opportunities, see Kodak or [http://www.dailytexanonline.com/blogs/the-update/2012/07/05/higgs-like-particle-texass-missed-opportunity](http://www.dailytexanonline.com/blogs/the-update/2012/07/05/higgs-like-particle-texass-missed-opportunity)

**Selected References:**

Suzuki, K., Wakisaka, S., & Fujii, N. (2012). Substitutional reality system: a novel experimental platform for experiencing alternative reality. Scientific reports, 2.

Takahashi, K., & Yamanaka, S. (2006). Induction of pluripotent stem cells from mouse embryonic and adult fibroblast cultures by defined factors. cell, 126(4), 663-676.

Takebe, T., Sekine, K., Enomura, M., Koike, H., Kimura, M., Ogaeri, T., ... & Taniguchi, H. (2013). Vascularized and functional human liver from an iPSC-derived organ bud transplant. Nature, 499(7459), 481-48

Velugotla, S., Pells, S., Mjoseng, H. K., Duffy, C. R., Smith, S., De Sousa, P., & Pethig, R. (2012). Dielectrophoresis based discrimination of human embryonic stem cells from differentiating derivatives. Biomicrofluidics, 6, 044113.

Vimal, R. (2009). Meanings attributed to the term'consciousness': an overview. Journal of Consciousness Studies, 16(5), 9-27.

Wickner, W., & Schekman, R. (2005). Protein translocation across biological membranes. science, 310(5753), 1452-1456.

Woolf, N. J., Priel, A., & Tuszynski, J. A. (2010). Nanoscale Components of Neurons: From Biomolecules to Nanodevices. In Nanoneuroscience (pp. 35-84). Springer Berlin Heidelberg.

Wu, G. (2005). *Nonlinearity and chaos in molecular vibrations*. Elsevier.
15